\newcommand{\feseven}{\mathrm{Fe^{7+}}}

\documentclass[manuscript]{aastex}
\usepackage{amsbsy}
\usepackage{amsfonts}
\usepackage{amssymb}

\usepackage{graphicx}
\usepackage{graphics}
\usepackage{bm}

\begin{document}


\title{Storage Ring Cross Section Measurements for Electron Impact Ionization of $\mathrm{\mathbf{Fe^{7+}}}$}

\author{M. Hahn\altaffilmark{1}, A. Becker\altaffilmark{2}, D. Bernhardt\altaffilmark{3}, M. Grieser\altaffilmark{2}, C. Krantz\altaffilmark{2},  M. Lestinsky\altaffilmark{4}, A. M\"{u}ller\altaffilmark{3}, O. Novotn\'{y}\altaffilmark{1,2}, R. Repnow\altaffilmark{2}, S. Schippers\altaffilmark{3}, K. Spruck\altaffilmark{3}, A. Wolf\altaffilmark{2}, and D. W. Savin\altaffilmark{1}}

\altaffiltext{1}{Columbia Astrophysics Laboratory, Columbia University, 550 West 120th Street, New York, NY 10027 USA}
\altaffiltext{2}{Max-Planck-Institut f\"{u}r Kernphysik, Saupfercheckweg 1, 69117 Heidelberg, Germany}
\altaffiltext{3}{Institut f\"{u}r Atom- und Molek\"{u}lphysik, Justus-Liebig-Universit\"{a}t Giessen, Leihgesterner Weg 217, 35392 Giessen, Germany}
\altaffiltext{4}{GSI Helmholtzzentrum f\"ur Schwerionenforschung, Planckstr. 1, 64291 Darmstadt, Germany }

\date{\today}
\begin{abstract}
	
	We have measured electron impact ionization (EII) for $\feseven$ from the ionization threshold up to 1200~eV. The measurements were performed using the TSR heavy ion storage ring. The ions were stored long enough prior to measurement to remove most metastables, resulting in a beam of 94\% ground state ions. Comparing with the previously recommended atomic data, we find that the \citet{Arnaud:ApJ:1992} cross section is up to about 40\% larger than our measurement, with the largest discrepancies below about 400~eV. The cross section of \citet{Dere:AA:2007} agrees to within 10\%, which is about the magnitude of the experimental uncertainties. The remaining discrepancies between measurement and the most recent theory are likely due to shortcomings in the theoretical treatment of the excitation-autoionization contribution. 

\end{abstract}

	
\maketitle
	
\section{Introduction}
	
	The spectrum of light emitted from astrophysical sources depends on the charge state distribution (CSD) of the gas. Interpreting these spectra therefore requires accurate calculations for the CSD \citep{Brickhouse:AIP:1996, Landi:AA:1999}. The CSD in collisionally ionized plasmas is determined by the balance between electron impact ionization (EII) and electron-ion recombination. Such plasmas are found in stellar coronae, supernova remnants, galaxies, and galaxy clusters. Accurate EII data are therefore needed in order to interpret observations of these objects. 
	
	Most EII data must come from theoretical calculations, as it is not practical to measure ionization for every astrophysically relevant ion. However, experiments provide important benchmarks for theory. A major limitation of most previous EII measurements has been that the ion beams used contained an unknown population of metastable ions. Since the cross section for ionization from a metastable level can differ signifcantly from that for ionization from the ground state, these data do not always provide a reliable benchmark for theory. 
	
	To rectify this problem, we have been performing EII measurements using an ion storage ring. In these experiments the ions are recirculated in the ion storage ring for several seconds before collecting data. This allows most metastable levels to radiatively relax to the ground state. The resulting measurements can thereby provide an unambiguous test for theoretical models. Our approach has been to measure EII for at least one ion in every astrophysically important isoelectronic sequence. These measurements can then be used to benchmark theoretical calculations, which can then be used to more accurately predict the cross sections for the unmeasured systems. A review of our work up to this point has been given by \citet{Hahn:JPCS:2014}. 
	
	 Here we focus on ionization of K-like ions, a system for which metastable contamination has been a problem in previous measurements. Experiments to measure ionization have been carried out for Sc$^{2+}$ \citep{Pindzola:PRA:1994}, Ti$^{3+}$ \citep{Falk:PRA:1983, VanZoest:JPhysB:2004}, and Mn$^{6+}$ \citep{Rejoub:PRA:2000}. \citet{Pindzola:PRA:1994} and \citet{Rejoub:PRA:2000} both note that there is a significant cross section below the ground state ionization threshold in their measurements. This is indicative of ionization from metastable levels, which have a lower ionization threshold energy compared to ground state ions. Neither of the Ti$^{3+}$ measurements show a clear ionization signal at energies below the ionization threshold, however the measurements by \citet{Falk:PRA:1983} and \citet{VanZoest:JPhysB:2004} are discrepant by up to 40\%. Part of this discrepancy may be attributable to metastables. 
	
We have performed storage ring measurements of for K-like $\feseven$ forming Ar-like Fe$^{8+}$. The single ionization cross section was measured from 70 to 1200~eV, which includes the direct ionization channels: 
\begin{equation}
\mathrm{e^{-} + Fe^{7+}} (2s^2\, 2p^6\, 3s^2\, 3p^6\, 3d) \rightarrow \left\{ \begin{array}{l} 
													\mathrm{Fe^{8+}} (2s^2\, 2p^6\, 3s^2\, 3p^6) + 2\mathrm{e^{-}} \\
													\mathrm{Fe^{8+}} (2s^2\, 2p^6\, 3s^2\, 3p^5\, 3d) + 2\mathrm{e^{-}} \\
													\mathrm{Fe^{8+}} (2s^2\, 2p^6\, 3s\, 3p^6\, 3d) + 2\mathrm{e^{-}} \\
													\mathrm{Fe^{8+}} (2s^2\, 2p^5\, 3s^2\, 3p^6\, 3d) + 2\mathrm{e^{-}} \\
													\mathrm{Fe^{8+}} (2s\, 2p^6\, 3s^2\, 3p^6\, 3d) + 2\mathrm{e^{-}} 
													\end{array} \right. .
\label{eq:transitions7}
\end{equation}	
The thresholds for ionization from the $3d$, $3p$, and $3s$ subshells are, 151.06, 201.4, and 241.2~eV, respectively \citep{NIST:2013}. Direct ionization of $2p$ and $2s$ electrons can occur above 933 and 1063~eV, respectively \citep{Kaastra:AAS:1993}. In these cases, however, direct ionization of electrons with a principal quantum number of $n=2$ produces a doubly excited state that stabilizes through autoionization with a probability of more than $90\%$ \citep{Kaastra:AAS:1993}. Thus, in most cases ionization of $n=2$ electrons leads to a net double ionization rather than single ionization. Excitation-autoionization (EA) is also predicted to contribute significantly to the single ionization cross section. The most important EA channels are expected to be $3p\rightarrow 5l$ and $3s \rightarrow 4l$ excitations followed by autoionization forming Fe$^{8+}$ \citep{Pindzola:NucFus:1987, Arnaud:ApJ:1992, Dere:AA:2007}. These channels are open starting from the ionization threshold. 

	The rest of this proposal is organized as follows: Section~\ref{sec:expan} describes the experimental method and the analysis, the influence of metastables on the measurement is discussed in Section~\ref{sec:meta}, our results are reported and discussed in Section~\ref{sec:results}, and Section~\ref{sec:sum} concludes.

\section{Experimental Method and Analysis}\label{sec:expan}

	Cross section measurements were carried out using the TSR heavy ion storage ring located at the Max-Planck-Institut f\"{u}r Kernphysik in Heidelberg, Germany \citep{Habs:NIMB:1989,Grieser:EJP:2012}. The procedures used here are basically the same as those we have described for previous EII measurements and are described in detail in \citet{Linkemann:fe15}, \citet{Hahn:mg7, Hahn:fe11, Hahn:fe12, Hahn:s12, Hahn:fe9, Hahn:fe13}, and \citet{Bernhardt:fe14}. Additionally, recombination measurements for Fe$^{7+}$ were performed at TSR by \citet{Schmidt:AA:2008}, using similar methods. Here we briefly review the procedure and provide some details relevant to the present measurement.
	
	A beam of $^{56}$Fe$^{7+}$ ions was injected into TSR with an energy of 82.1~MeV. In order to efficiently fill the ring with ions, the ions were introduced in seven pulses separated in time by 1.47~s for most of our work and followed then by a 30-s cooling period (see below). As will be discussed below, in order to quantify the effect of metastable levels in the beam we also used a shorter injection cycle with only a single injection pulse for a few experimental runs. Following injection, the ions circulated around the ring where they were merged with an electron beam, called the Cooler, with a very small energy spread. Initially the velocity of the electrons was set to match the average ion velocity so that elastic electron-ion collisions reduced the energy spread of the ion beam. This process is known as electron cooling \citep{Poth:PhysRep:1990}. For most of our work this cooling period lasted 30~s, but for a few runs we used a shorter timing with only 1.5~s of cooling. The reason for using a long cooling time is to allow the metastable levels to radiatively decay before data collection starts. The possible influence of metastable levels in the beam is discussed in more detail below in Section~\ref{sec:meta}. 
	
	Following the initial cooling period, the energy of the electron beam was varied so that the ionization cross section could be measured at different energies. Ionized products from electron-ion collisions in the Cooler electron beam were separated from the parent beam by a dipole magnet downstream of the interaction region and directed onto a particle counting detector. Following each measurement at the variable collision energy, a reference measurement was performed at a fixed reference energy. This allowed us to determine the background rate due to electron stripping in residual gas collisions. For measurement energies up to 780~eV the reference energy was chosen to be 78~eV, which is well below the ionization threshold. The signal at this reference energy consisted only of background from stripping on the residual gas in TSR. However, due to the limited dynamic range of the electron beam power supply,  at energies above 780~eV the reference point was above the ionization threshold and the reference count rate had to be corrected for counts due to EII at the reference energy. This correction has been described in detail in our previous work \citep[e.g.,][]{Hahn:fe12}. 
	
	The EII cross section was calculated by subtracting the background rate from the measured ionization count rate and normalizing by the stored ion number and the electron density \citep[e.g.,][]{Hahn:fe11}. The uncertainties in the measurement are summarized in Table~\ref{table:error}. Here and throughout all uncertainties are given at an estimated $1\sigma$ level. The detector efficiency introduces an uncertainty of about 3\% \citep{Rinn:RSI:1982}. There is also a 3\% uncertainty in the electron density \citep{Lestinsky:ApJ:2009}. Unlike most of our previous measurements, the beam was not continuously cooled throughout the measurement. As a result, warming caused the ion beam to expand from 1~mm full width at half maximum (FWHM) immediately following the cooling step to 2~mm FWHM by the end of the measurement cycle. The detector had a sensitive area with a diameter of 10~mm, thereby enabling us to still collect the full Fe$^{8+}$ product ion beam. The method of \citet{Lampert:PRA:1996} was used to correct for the energy spread from the toroidal merging and demerging sections where the ion and electron beams are not co-linear. 
		
		The largest source of experimental error is from the ion current measurement. This current is measured using a beam profile monitor \citep[BPM;][]{Hochadel:NIMA:1994}, which has a calibration that fluctuates over time. We used a DC transformer \citep{Unser:IEEE:1981} to calibrate the BPM several times during the experiment. The DC transformer has a stable calibration, but is not sensitive to the low currents of $\sim 1$~$\mathrm{\mu A}$ present during measurement. For this calibration the ring was filled with a high current of ions, up to 48~$\mathrm{\mu A}$. We estimate that the uncertainty in the calibration of the ion current leads to a 10\% systematic error in the final cross section. Because this uncertainty determines the normalization of the measurement, it affects the magnitude of the cross section but does not distort the shape as a function of energy. 
	
	Corrections were also performed to account for energy dependent changes in the pressure, using the method described in \citet{Hahn:mg7}. The correction was small, $\approx 1\%$. However, as the correction can only be applied to data where the reference point is below threshold, measurements at collision energies above 780~eV were not corrected for this slight distortion and therefore have an estimated additional 1\% uncertainty not present in the lower energy data. 
	
\section{Metastables}\label{sec:meta}

	For Fe$^{7+}$ there are two metastable levels with lifetimes long enough to be of particular concern in this work. Calculations for the radiative transition rates and lifetimes for Fe$^{7+}$ have been performed by \citet{Tayal:ApJ:2011} and \citet{DelZanna:AA:2009}. For simplicity, we will quote the lifetimes derived from the calculations of \citet{DelZanna:AA:2009}, which are tabulated in the CHIANTI atomic database \citep{Landi:ApJ:2013}.  The Fe$^{7+}$ ground state is the $3p^6\,3d\;^{2}D_{3/2}$ level. Lying 0.23~eV above the ground level is the $3p^6\,3d\;^{2}D_{5/2}$ level, which is metastable with a lifetime of about 12.7~s. Metastables from this level can be removed by allowing for a reasonable delay between injection and measurement. The other problematic metastable is the $3p^5\,3d^2\,(^{3}F)^{4}G_{11/2}$ level lying 51.36~eV above the ground level and having an estimated lifetime of 2867~s. This lifetime is too long to expect radiative transitions to remove these metastables from the beam. 
	
\begin{figure}
\centering \includegraphics[width=0.9\textwidth]{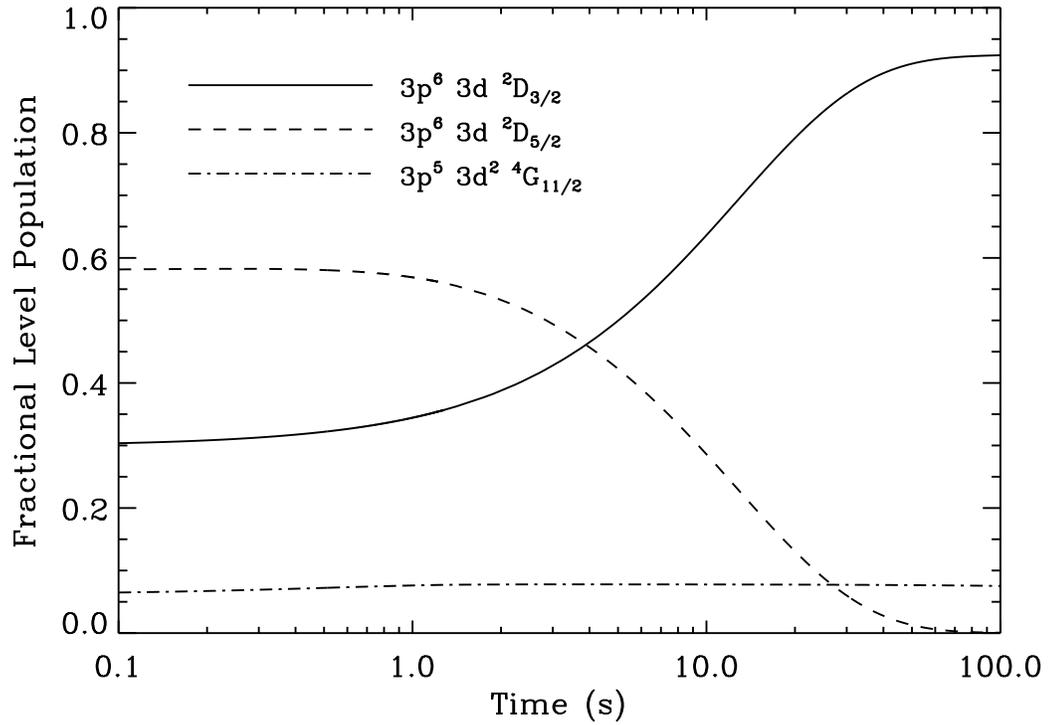}
\caption{\label{fig:fe7metapop} Modeled level populations for the ground level $3p^6\,3d\;^{2}D_{3/2}$ and the two long-lived metastable levels $3p^6\,3d\;^{2}D_{5/2}$ and $3p^5\,3d^2\;^{4}G_{11/2}$. The model starts from a Boltzmann distribution with $k_{\mathrm{B}}T=750$~eV and solves for the level population evolution due to radiative transitions.
}
\end{figure}

	In order to estimate the metastable fraction in the ion beam, we have used the procedure described by \citet{Lestinsky:ApJ:2012} to evolve the level populations in time, based on the radiative transition rates. For the radiative transition data, we used the rates from the CHIANTI database, which includes 104 levels from the ground state up to an excitation energy of 132.66~eV \citep{DelZanna:AA:2009, Landi:ApJ:2013}. The initial populations for this calculation were determined using a Boltzmann distribution with $k_{\mathrm{B}}T_{\mathrm{e}} = 100$~eV, where $k_{\mathrm{B}}$ is the Boltzmann constant and $T_{\mathrm{e}}$ is the electron temperature. This corresponds to an ion energy of about 10~MeV and is an estimate of the effective temperature of the electrons, as seen by the ions, in the gas stripper when the Fe$^{7+}$ ions are formed. Figure~\ref{fig:fe7metapop} shows the fractional population of the ground level and the two metastable levels discussed above. If the only factor affecting the relative populations were radiative decay, this calculation shows that the population of the metastable $^{2}D_{5/2}$ level would be $\approx4\%$ with the longer 30-s cooling time and $\approx37\%$ with the shorter 1.5-s cooling time. In either scenario the population of the long-lived $^{4}G_{11/2}$ level would be $\approx8\%$ throughout the measurement. 
	
\begin{figure}
\centering \includegraphics[width=0.9\textwidth]{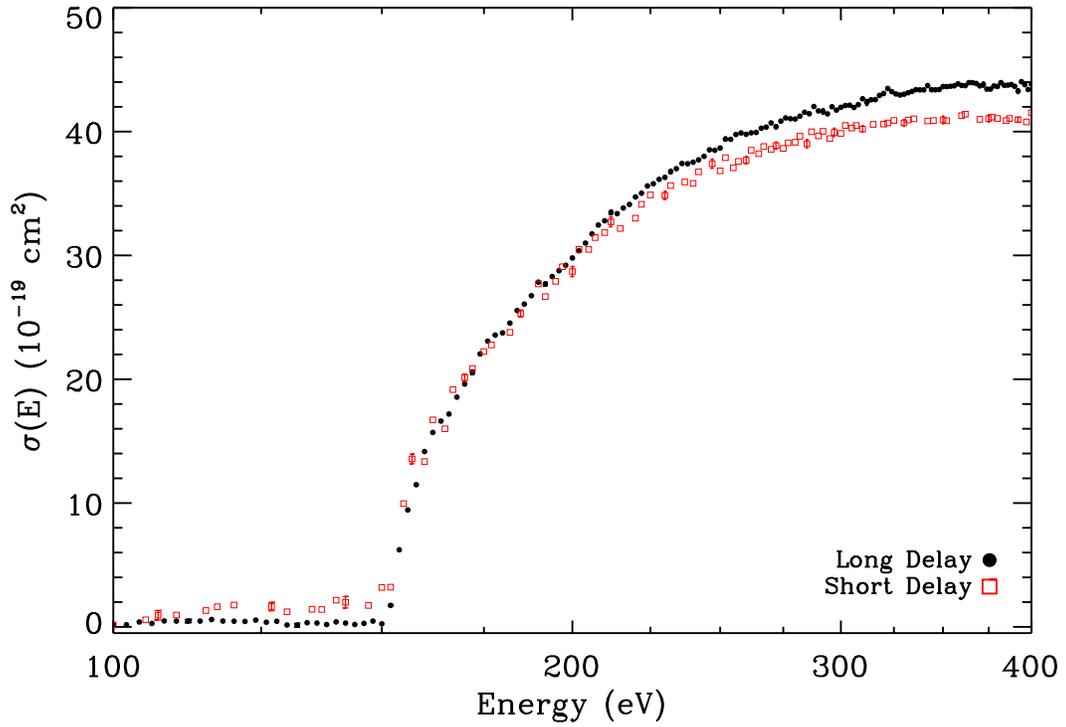}
\caption{\label{fig:fe7metacross} The Fe$^{7+}$ ionization cross section derived from measurements with a long delay time of $30$~s and a short delay time of $1.5$~s. In the short-delay results, the nonzero cross section below $\approx 150$~eV is due to the ionization of $3p^5\,3d^2\;^{4}G_{11/2}$ metastable ions in the beam. 
}
\end{figure}

	
	Figure~\ref{fig:fe7metacross} compares the cross sections measured with both the long and short delay times between injection and data collection. For the short delay, there is clearly a significant cross section for ionization starting about 50~eV below the ground state ionization threshold of 151.06~eV. We attribute this cross section to the presence of the $3p^5\,3d^2\,(^{3}F)^{4}G_{11/2}$, metastable level. In the data collected with the long delay time, the below threshold signal has largely disappeared. If we assume that the decay of the metastables is exponential then, for a given energy the ratio of the cross section with the long delay time to that with the short delay time is $\sigma_{\mathrm{I}}^{\mathrm{long}}/\sigma_{\mathrm{I}}^{\mathrm{short}} = \mathrm{e}^{-\Delta t/\tau}$, where $\tau$ is the metastable lifetime in the beam and $\Delta t$ is the timing difference. Using the data in the range $105$--$151$~eV, we find $\tau=24 \pm 10$~s. This is considerably shorter than the theoretically predicted radiative decay rate; but our inferred rate does not necessarily imply that the calculated rates are in error. One alternative explanation is that the cross section for stripping or electron capture from the residual gas is larger for the metastables than for the ground state, and so metastables are preferentially removed from the ion beam. Collisions might also excite metastables to levels that decay to the ground state. Alternatively, the shorter than expected lifetime of the $^{4}G_{11/2}$ level could be due to quenching in the magnetic fields of the storage ring \citep{Grumer:PRA:2013}. Regardless of the cause, the observed large effective metastable decay rate results in a fraction of $^{4}G_{11/2}$ metastable levels in the ion beam following the long delay estimated to be about $2\pm1$\%. 
	
	In both the short and the long delay-time data, the cross sections also contain some contribution from the $^{2}D_{5/2}$ metastable level in the ground configuration. Unfortunately, the additional contribution to the cross section from this metastable level cannot be resolved because of the small energy difference from the ground level. 
	
	The total metastable fraction in our data, using the long delay time, is estimated at $\approx6\%$. Although metastables are likely still present in our ion beam during measurement, the storage ring experiment has significantly reduced their impact compared to previous work. The most closely related previous measurement is the crossed beams measurement of the isoelectronic Mn$^{6+}$ by \citet{Rejoub:PRA:2000}. In those results a significant cross section can be seen below the ground state ionization threshold, whereas Figure~\ref{fig:fe7metacross} shows that with a long initial delay time, this contribution is reduced nearly to zero in our storage ring experiment. 
	
	To quantify the systematic uncertainty in our results, we consider the relative size of the EII cross section from the metastable and ground state ions. Figure~\ref{fig:fe7metacross} shows that above threshold the cross section with the long delay time is slightly larger than with the short delay time, though within our 10\% systematic uncertainty. This suggests that the metastable EII cross section is similar to or slightly smaller than the ground level cross section. To give an upper bound on the uncertainty from the metastables, we can conservatively estimate the cross section for EII from the metastable level to be between zero and twice the cross section from the ground level. Then the systematic uncertainty in the cross section due to metastables is $\lesssim6\%$, i.e., approximately equal to the metastable fraction. 

\section{Results and Discussion} \label{sec:results}

\subsection{Cross Section}\label{subsec:cross}

\begin{figure}
\centering \includegraphics[width=0.9\textwidth]{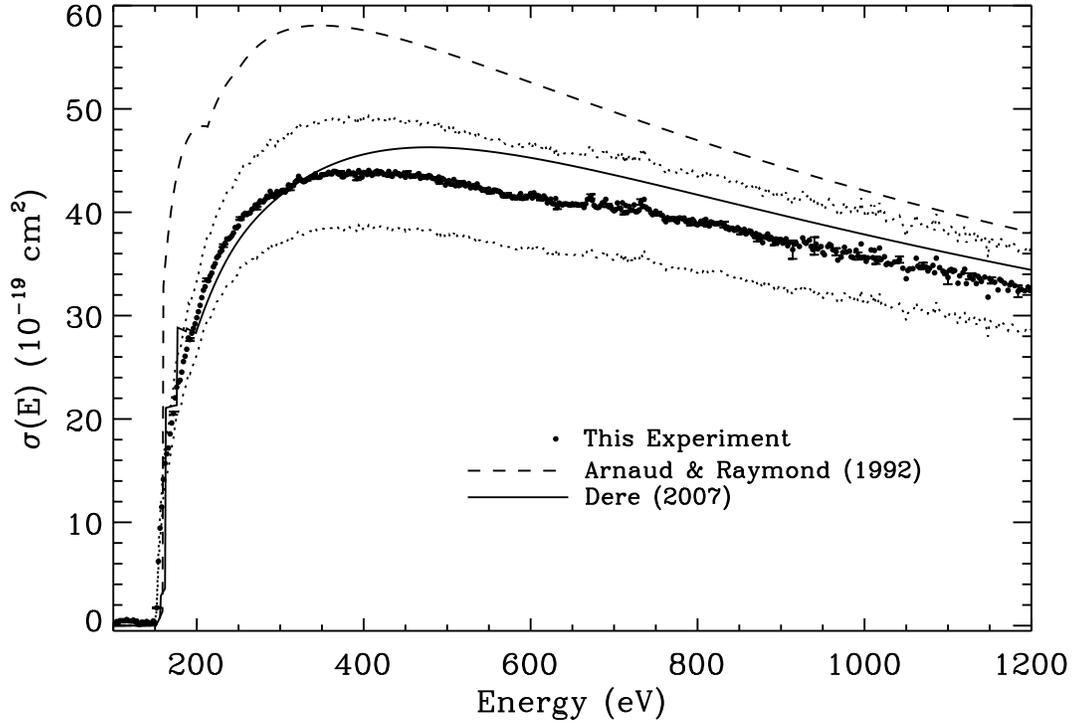}
\caption{\label{fig:fe7cross} EII cross section for $\feseven$ forming Fe$^{8+}$ (circles). The dotted curves illustrate the $1\sigma$ systematic uncertainties, mainly due to the ion current measurement and metastable contamination. Statistical uncertainties are indicated by the error bars on selected points, which are in some cases smaller than the symbols. The dashed curve shows the recommended cross section from \citet{Arnaud:ApJ:1992} and the solid curve illustrates the cross section calculated by \citet{Dere:AA:2007}. 
}
\end{figure}

	Figure~\ref{fig:fe7cross} shows our measured ionization cross section for $\feseven$ forming Fe$^{8+}$. These data are also available in the electronic edition of this journal as a table following the format of Table~\ref{table:fe7cross}. In Figure~\ref{fig:fe7cross} the filled circles show the measured cross section and the dotted curves indicate the $1\sigma$ systematic uncertainty, which mainly arises from the ion current measurement and the metastable contamination. The error bars on selected points illustrate the $1\sigma$ statistical errors, which are about 1\%. 
	
	The dashed line in Figure~\ref{fig:fe7cross} shows the cross section recommended by \citet{Arnaud:ApJ:1992}, which is based on the calculations of \citet{Pindzola:NucFus:1987}. Near threshold this cross section is up to twice as large as the measurement and at energies higher than that where the cross section peaks the theory remains 20--30\% larger than our experimental results. Part of the difference is that these calculations overestimate the EA contribution near the ionization threshold. The solid line in the figure indicates the cross section calculated by \citet{Dere:AA:2007}. We find reasonably good agreement with these calculations, with differences of only $\sim10\%$. The \citet{Dere:AA:2007} calculations include EA from $3p^6\,3d \rightarrow 3p^5\,3d\, nl$ levels for $n=4,5$ and $6$. Thus, they omit excitations of the $3s$ electron leading to EA, which were predicted by \citet{Pindzola:NucFus:1987} to be important. That our results agree with the \citet{Dere:AA:2007} calculations, which exclude the $3s$ EA channel, implies that either the $3s$ EA is actually smaller than predicted by \citet{Pindzola:NucFus:1987}, or that \citet{Dere:AA:2007} overestimates the EA arising from the $3p$ electrons, and this error is compensated for by the neglect of the $3s$ EA channel in that work. 
	
	One other feature visible in the measured cross section, is an increase at about 600~eV. This energy corresponds to excitation from the $n=2$ level, and the increased cross section here is attributed to EA from such excitations. These EA channels were not considered in either \citet{Arnaud:ApJ:1992} or \citet{Dere:AA:2007}. However, this does not lead to a large error because the increase is small, only a few percent. Near this increase, we also see some resonances. We attribute these resonances to dielectronic capture into excited levels that relax by double ionization, thereby leading to a net single ionization \citep{Muller:AAMOP:2008}.

\subsection{Rate Coefficients}\label{subsec:rate}

\begin{figure}
\centering \includegraphics[width=0.9\textwidth]{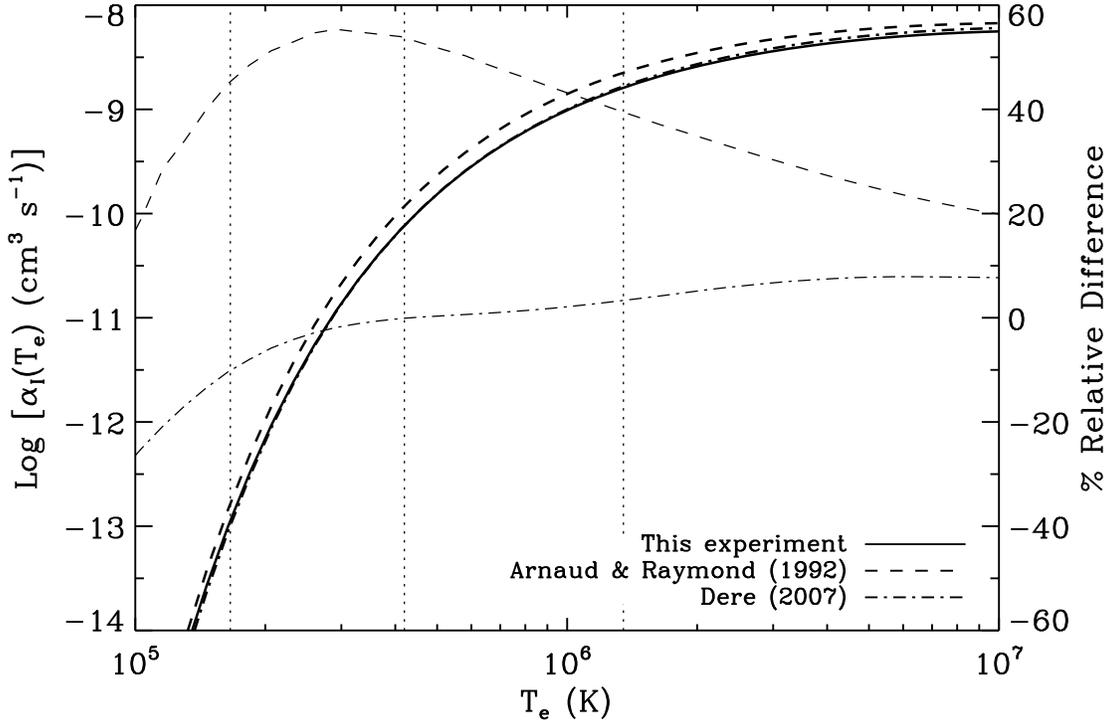}
\caption{\label{fig:fe7rate} Thick lines show the plasma rate coefficient for $\feseven$ forming Fe$^{8+}$, which can be read off the left axis. The thick solid curve indicates the experimental results, which are compared to the theoretical results of \citet[][thick dashed curve]{Arnaud:ApJ:1992} and \citet[][thick dash-dotted curve]{Dere:AA:2007}. The relative difference between these and the present results, (theory-experiment)/experiment, are shown by thin lines, with values read off the right axis. Dotted vertical lines indicate the temperature range where $\feseven$ is $>1\%$ abundant in collisional ionization equilibrium with the center line showing the temperature of peak $\feseven$ abundance \citep{Bryans:ApJ:2009}. 
}
\end{figure}

	We have derived plasma ionization rate coeffieicnts $\alpha_{\mathrm{I}}(T_{\mathrm{e}})$ from the measured cross sections, as described in \citet{Hahn:fe11}. Figure~\ref{fig:fe7rate} shows the inferred plasma rate coefficient. Fe$^{7+}$ is greater than 1\% abundant in collisional ionization equilibrium from $1.7 \times 10^{5}$~K to $1.4 \times 10^{6}$~K, with peak abundance at $4.2 \times 10^{5}$~K \citep{Bryans:ApJ:2009}. These temperatures are indicated on the plot by vertical dotted lines. 
	
	The rate coefficients inferred from our measurement are compared to those of \citet{Arnaud:ApJ:1992} and \citet{Dere:AA:2007}, indicated by the thick dashed and dash-dotted lines, respectively. The relative differences, theory$-$experiment/experiment, are illustrated by the thin curves, which can be read off the right axis. It can be seen that the \citet{Arnaud:ApJ:1992} rate coefficient is $\gtrsim 40\%$ larger than the measured rate coefficient over the temperature range at which Fe$^{7+}$ is abundant. The \citet{Dere:AA:2007} rate coefficient agrees to within 10\% with our results over these temperatures.

	 Table~\ref{table:coeff} provides coefficients for a polynomial fit to a scaled rate coefficient 
\begin{equation}	 
\rho(x)=10^{-6}\sum_{i}{a_i x^i}, 
\label{eq:rhosum}
\end{equation}
which can be used to reproduce our results. The rate coefficient $\alpha_{\mathrm{I}}(T_{\mathrm{e}})$ is related to the scaled rate coefficient $\rho$ by \citep{Dere:AA:2007}: 
\begin{equation}
\alpha_{\mathrm{I}}(T_{\mathrm{e}}) = t^{-1/2}E_0^{-3/2}E_{1}(1/t)\rho(x), 
\label{eq:invscalerate}
\end{equation}
where $E_{1}(1/t)$ is the first exponential integral and $t=k_{\mathrm{B}}T_{\mathrm{e}}/ E_0$ with $E_0=151.06$~eV the ionization threshold. The scaled temperature $x$ is given by
\begin{equation}
x = 1 - \frac{\ln 2}{\ln(t+2)}
\label{eq:invx}
\end{equation}
and by inverting $T_{\mathrm{e}}$ can be obtained from $x$: 
\begin{equation}
T_{\mathrm{e}} = \frac{E_0}{k_{\mathrm{B}}}\left[\exp\left(\frac{\ln 2}{1-x} \right) - 2 \right]. 
\label{eq:invscaletemp}
\end{equation}
The experimental rate coefficents are reproduced to better than 1\% accuracy for $T_{\mathrm{e}} = 8\times10^{4}$ -- $1\times10^{8}$~K.  	

\section{Summary}\label{sec:sum}	

	We have measured EII for Fe$^{7+}$ forming Fe$^{8+}$ using an ion storage ring. This ion has several long-lived metastable levels, but by using a long storage time we were able to reduce the metastable contamination in the ion beam to $\approx 6\%$ during measurements. The result is a nearly pure measurement of the ground state EII cross section. We find that this cross section is significantly smaller than the previously recommended cross section of \citet{Arnaud:ApJ:1992}. It is in reasonable agreement with the calculations of \citet{Dere:AA:2007}; however, some differences in the shape of the cross section remain. The discrepancies are likely due to the treatment of EA in the theoretical calculations.	
	
\begin{acknowledgments}
	We appreciate the efficient support by the MPIK accelerator and TSR groups during the beamtime. This work was supported in part by the NASA Astronomy and Physics Research and Analysis program and the NASA Solar Heliospheric Physics program. We also acknowledge financial support by the Max Planck Society, Germany, and from Deutsche Forschungsgemeinschaft (contract no. Schi 378/8-1).  
\end{acknowledgments}

\begin{deluxetable}{lc}
\tablecolumns{2}
\tablewidth{0pt}
\tablecaption{Sources of Uncertainty.
\label{table:error}}
\tablehead{
	\colhead{Source} & 
	\colhead{Estimated $1\sigma$ Uncertainty}
}
\startdata
Counting statistics 										& 1\% 	\\
Detector efficiency											& 3\%		\\
Electron density 												& 3\% 	\\
Ion current measurement 								& 10\%	\\
Metastable contamination 								& 6\%		\\
Pressure fluctuations\tablenotemark{1} 	& 1\% \\
\hline
Quadrature sum 												&  12\%		\\
\enddata
\tablenotetext{1}{The uncertainty from the pressure fluctuations applies to energies above $780$~eV where the data could not be corrected. For lower energy data, the uncertainty in the correction itself is a fraction of a percent, and so considered negligible.}
\end{deluxetable}

\clearpage

\begin{deluxetable}{lll}
\tablecolumns{3}
\tablewidth{0pc}
\tablecaption{$\feseven$ Ionization Cross Section.
\label{table:fe7cross}}
\tablehead{
	\colhead{$E$~(eV)} & 
	\colhead{$\sigma_{\mathrm{I}}$~(cm$^2$)} &
	\colhead{Statistical Error}
}
\startdata
200 & 2.979E-018 & 1.414E-020 \\
400 & 4.384E-018 & 1.656E-020 \\
600 & 4.182E-018 & 1.764E-020 \\
800 & 3.878E-018 & 2.219E-020 \\
1000& 3.565E-018 & 7.664E-020 \\
1200& 3.263E-018 & 2.879E-020 
\enddata
\tablecomments{There is a systematic uncertainty of 12\% in the cross section (see the text). Table~\ref{table:fe7cross} is published in its entirety in the electronic edition of this journal.}
\end{deluxetable}

\begin{deluxetable}{llllll}
\tablecolumns{4}
\tablewidth{0pc}
\tablecaption{Sixth-order Polynomial Fitting Parameters to Reproduce the Scaled Single Ionization Rate Coefficient $\rho = 10^{-6}\sum_{i=0}^{i=5}{a_{i}x^{i}}$~$\mathrm{cm^{3}\,s^{-1}\,eV^{3/2}}$, see Equations (\ref{eq:invscalerate}) and (\ref{eq:invscaletemp}).
\label{table:coeff}}
\tablehead{
	\colhead{$i$} & 
	\colhead{$a_{i}$}
}
\startdata 
0 & \phs43.4136 \\
1 & \phn-329.950 \\
2 & \phs1987.92 \\
3 & \phn-6312.25 \\
4 & \phs10734.7 \\
5 & \phn-9190.33 \\
6 & \phs3097.61 
\enddata
\end{deluxetable}

\bibliography{EII}

\end{document}